\documentclass[manuscript,screen,nonacm]{acmart}
\AtBeginDocument{%
  }

\begin{document}

\title{Identifying Likely-Reputable Blockchain Projects on Ethereum}

\author{Cyrus Malik}
\email{cyrus.malik.20@um.edu.mt}
\affiliation{%
  \institution{Department of Computer Science, University of Malta}
  \country{Malta}
}

\author{Josef Bajada}
\affiliation{%
  \institution{Department of Artificial Intelligence, University of Malta}
  \country{Malta}}
\email{josef.bajada@um.edu.mt}

\author{Joshua Ellul}
\affiliation{%
  \institution{Centre for DLT \& Department of Computer Science, University of Malta}
  \country{Malta}
}

\renewcommand{\shortauthors}{Malik et al.}

\begin{abstract}
  Identifying reputable Ethereum projects remains a critical challenge within the expanding blockchain ecosystem. The ability to distinguish between legitimate initiatives and potentially fraudulent schemes is non-trivial. This work presents a systematic approach that integrates multiple data sources with advanced analytics to evaluate credibility, transparency, and overall trustworthiness. The methodology applies machine learning techniques to analyse transaction histories on the Ethereum blockchain.

The study classifies accounts based on a dataset comprising 2,179 entities linked to illicit activities and 3,977 associated with reputable projects. Using the LightGBM algorithm, the approach achieves an average accuracy of 0.984 ($\pm 0.003$) and an average AUC of 0.999, validated through 10-fold cross-validation. Key influential factors include time differences between transactions and received\_tnx.

The proposed methodology provides a robust mechanism for identifying reputable Ethereum projects, fostering a more secure and transparent investment environment. By equipping stakeholders with data-driven insights, this research enables more informed decision-making, risk mitigation, and the promotion of legitimate blockchain initiatives. Furthermore, it lays the foundation for future advancements in trust assessment methodologies, contributing to the continued development and maturity of the Ethereum ecosystem.
\end{abstract}


\maketitle

\section{Introduction}
The Ethereum blockchain has witnessed instances where certain entities or users exploit its pseudonymity to engage in illicit activities. This raises the fundamental question of whether it is possible to systematically differentiate reputable projects from those that may not be. While existing research has primarily focused on detecting fraudulent activities—such as scams, Ponzi schemes, and network anomalies—these efforts remain centered on identifying and flagging illicit behavior rather than providing a holistic assessment of a project’s overall reputability.

Several studies have explored the detection of illicit activities on the Ethereum blockchain \cite{farrugia2020detection}, the identification of Ponzi schemes \cite{yu2021ponzi}, for anti-money laundering \cite{vassallo2021application} and anomaly detection within the network \cite{sayadi2019anomaly}. While these contributions enhance our understanding of fraudulent behavior, they do not directly address the broader issue of evaluating whether a project itself is reputable.

Given the growing number of Ethereum-based initiatives, the need for a systematic approach to assessing project reputability becomes increasingly evident. Distinguishing between legitimate and potentially deceptive ventures requires a dedicated methodology that extends beyond merely detecting illicit activity. By establishing such an approach, stakeholders, including investors, developers, and regulators can make more informed decisions, mitigate risks associated with unreliable projects, and foster a more secure and transparent investment landscape within the Ethereum ecosystem.

This research aims to identify projects that are likely to be reputable by comparing them against a model comprised of data associated with a list of reputable projects from a source deemed to be trust-worthy. We therefore, define the following project aim to: develop a comprehensive methodology for identifying likely‐reputable Ethereum Blockchain based projects using transactional data and machine learning techniques.

To achieve the stated objectives, a structured approach is adopted, with each stage building upon the previous one. The first step involves developing the foundational architecture for the final deliverable tool, ensuring flexibility for future enhancements and refinements. The next phase focuses on constructing a robust dataset tailored for the classification of Ethereum accounts, leveraging a supervised LightGBM model. Finally, the methodology is validated through a comprehensive evaluation of the model’s performance in distinguishing between likely reputable and illicit addresses.

\section{Background and Related Work}
In the context of evaluating reputable projects on the Ethereum blockchain, \cite{alzahrani2020predicting} explore the application of machine learning techniques to predict the success of Initial Coin Offerings (ICOs). ICOs served as a fundraising mechanism for blockchain initiatives. However, their high failure rate and the prevalence of fraudulent schemes pose significant challenges for both investors and regulators. The study by \citeauthor{alzahrani2020predicting} introduces a machine learning-based approach that incorporates various features, including social media sentiment, project characteristics, and market indicators to assess the likelihood of an ICO’s success or failure. While the primary focus of their paper is on ICOs, the methodology aligns with the broader objective of identifying reputable Ethereum-based projects through machine learning. By leveraging transactional data and key project-related attributes, machine learning models provide a systematic means to assess credibility and predict project viability, contributing to a more transparent and informed investment landscape within the Ethereum ecosystem.

``Ethereum Smart Contracts Analysis: Detecting Ponzi Schemes'' by \citeauthor{zaremba2020ethereum} \cite{zaremba2020ethereum}, examines Ethereum smart contracts to identify Ponzi schemes. While the primary focus is on detecting fraudulent activity, the underlying methodology is relevant to assessing project reputability. The authors propose a machine learning-driven approach that analyses smart contract transaction data to uncover patterns indicative of Ponzi schemes. Key features extracted from transaction data include transaction frequency, contract balance variations, and interactions involving multiple contracts. These attributes are then leveraged to train a machine learning model capable of classifying smart contracts as either Ponzi or non-Ponzi. Although the study is specific to Ponzi scheme detection, its approach contributes to the broader effort of evaluating the legitimacy of Ethereum-based projects through data-driven analysis.

\subsection{LightGBM Model background}
LightGBM, short for Light Gradient Boosting Machine, is an optimised gradient boosting decision tree algorithm designed for efficiency and scalability. It builds upon the well-established decision tree methodology, leveraging the gradient boosting framework to enhance predictive performance. A fundamental concept underlying LightGBM is bootstrap aggregating, or bagging, which generates multiple classifiers from the training dataset \cite{breiman1996bagging}. Bagging involves sampling with replacement, producing diverse training subsets that vary in composition—some instances may appear multiple times, while others may be omitted. This iterative process is repeated for a predefined number of trials, resulting in a collection of training sets that improve model robustness and generalisation.

The next layer in this hierarchical approach is random forests, which refine the construction of decision and regression trees by incorporating additional randomness \cite{liaw2002classification}. Unlike standard bagging, random forests introduce variability at multiple levels. Each tree is built from a different bootstrap sample, and at each node, only a randomly selected subset of features is considered for the best split. This stochastic approach enhances model robustness and mitigates overfitting, a common challenge in machine learning.

The algorithm can be structured into three key stages for both classification and regression. First, bootstrap samples are drawn from the training dataset to create diverse subsets. Next, an unpruned decision tree is constructed for each sample. However, rather than selecting the optimal split from all available predictors, a random subset of predictors is chosen, and the best split is identified from this subset. When the subset size equals the total number of predictors, the process effectively reduces to bagging. Finally, predictions from all trees are aggregated using majority voting for classification tasks and averaging for regression---yielding the final output.

Boosting is a sequential training technique that is integrated into random forest approaches. It involves training a series of base classifiers in a sequential manner, where the error function is influenced by the performance of all the preceding models in the sequence \cite{bishop2006pattern}. Instances that were misclassified by previous classifiers are assigned higher weights, which affect subsequent classifiers.

Each base classifier is trained on a weighted form of the training set, with weights determined by the prior base classifier. After training, the weak classifiers are combined to form the final classifier known as the Strong classifier. AdaBoost, short for Adaptive Boosting, is a widely used implementation of the boosting algorithm in statistical computing languages \cite{adaboost}.

An enhancement to the boosting algorithm is\textit{ Gradient Boosting}, which applies gradient descent algorithms with boosting techniques to minimise total error iteratively by optimising parameters in the descending direction. Challenges such as local minima and plateaus are encountered during optimisation, where the optimal solution may appear to be reached \cite{wang2004improved, mitchell2006discipline}. The choice of the loss function depends on the scenario, as long as it is differentiable. The learning rate is an important parameter that affects the step size. A low learning rate requires more iterations to find the global minima, while a high learning rate increases the risk of surpassing local minima \cite{learning_rate}.

LightGBM's performance and speed can be attributed to two methodologies employed which are GOSS (Gradient-based One-Side Sampling) and EFB (Exclusive Feature Bundling) \cite{ke2017lightgbm}.

GOSS is a sampling strategy that focuses on training instances with larger gradients. It samples a subset of instances based on their gradient values, giving more weight to those with larger gradients. By doing so, it emphasises instances that contribute more to the overall loss function, leading to faster convergence during training. This approach helps reduce the computational cost of training the model while maintaining or even improving its predictive performance.

EFB, on the other hand, is a feature bundling technique that aims to enhance the efficiency of LightGBM in handling high-dimensional data. It groups exclusive features together based on their importance and correlation, creating a compact representation of the original features. This bundling reduces the memory footprint and computational overhead of the algorithm while preserving the information contained in the original features. It can be particularly beneficial when dealing with datasets that have a large number of features, improving the training speed and memory usage of LightGBM.

\section{Methodology}
\subsection{Data Acquisition}
To identify likely reputable projects, a list of smart contract addresses associated with reputable projects was collected from CoinGecko.\footnote{\url{https://www.coingecko.com/en/coins/ethereum}} A total of 4,060 smart contracts were obtained, representing projects that were classified as likely reputable --- indeed CoinGecko is not a definitive source of whether a project is definitively reputable, yet it provides an initial base towards some level of reputability given the due diligence required to be listed.~\footnote{\url{https://www.coingecko.com/en/listing_terms}}

Following data collection, feature extraction was performed using the same methodology as Farrugia et al., to the extent permitted by available data. Key features—such as the number of received and sent transactions—were derived using the Etherscan API\footnote{\url{https://etherscan.io/apis}}, which provides access to the most recent 10,000 transactions executed by either standard Ethereum accounts or ERC20-compliant accounts. The ERC20 standard defines a common set of rules and guidelines for fungible tokens on the Ethereum blockchain, ensuring interoperability across decentralised applications.

The \emph{Ethereum Illicit Accounts dataset} \cite{farrugia2020detection} used by Farrugia et al. consists of 42 features per account. Most of these features were recreated for the list of projects from CoinGecko that we identify to be likely-reputable.A subset of the extracted features showed similarities to those obtained in related research utilising graph analysis \cite{chen2020understanding}. The process of feature extraction was conducted in conjunction with the Etherscan transaction request API, and by performing a BigQuery (a web service provided by Google Cloud Platform that offers a fully managed and serverless data warehouse for analysing large datasets) on the \emph{Ethereum Blockchain} dataset provided by Google\footnote{\url{https://www.kaggle.com/datasets/bigquery/ethereum-blockchain}}. Most of the features were generated by using the Etherscan API however, for features such as total Ether sent and total Ether received, the big query dataset was used as it encompasses over a larger time frame. \\
Moreover, since the dataset provided by \cite{farrugia2020detection} only considers transactions for illicit and normal accounts till June 2019, the same date range was used when collecting data for the likely-reputable Ethereum accounts data-set's set of features to avoid bias in data collection. \\
Finally, some features were not included in the final dataset as the Etherscan transaction request API returned no value for more than 76\% of the smart contract addresses from CoinGecko.
The final dataset contains 6,156 instances where 2,179 of these are illicit  taken from \cite{farrugia2020detection} and 3,978 likely-reputable projects from CoinGecko. Moreover, the dataset also consists of the 38 features from the 42 features collected in \cite{farrugia2020detection}. 

\subsection{Data Visualisation and ML models}
To visualise the final dataset, a technique called t-Distributed Stochastic Neighbor Embedding (t-SNE) was employed, which reduces the dimensionality of the dataset containing 38 dimensions. The t-SNE is a non-linear method which allows visualisation of high dimensional data \cite{van2008visualizing}. This is conducted by mapping each data point to a position in a 2 or 3 dimensional space. Specifically, a conversion is carried out from high-dimensional euclidean distances present among the data points to conditional probabilities representing similarities \cite{arnoldi1951principle}. Thus, given a high dimensional dataset $X = x_{1},x_{2},x_{3},..,,x_{n}$ a conversion takes place such that the resultant dimensionality $Y = y_{1},y_{2},y_{3},..,,y_{n}$ may be visually represented as a scatter plot. \\
The main objective is to maintain the structural integrity within the high-dimensional space while transforming it into a lower-dimensional space, thus ensuring the discovery of any possible valuable insights. These insights would correspond to any relative similarities present in the accounts obtained. We implement and provide the resultant 2D and 3D t-SNE plots with respect to the account class label.

This was then followed by the implementation of the LightGBM classification model involving a sequential three-step process; (i) declaring the model along with all the respective parameters to be considered, ii) training the classifier on the training set and iii) predicting on the test data using the model trained in the previous step. Cross-validation was used to asses the model's performance. 

We assessed the following three main LightGBM parameters using grid-search; i) learning rate, ii) the number of estimators (n\_estimators) and iii) the maximum depth (max\_depth).

The best parameters were determined using the Receiver Operating Characteristic (ROC) curve which presents us with the area under the curve (AUC) measure \cite{bradley1997use}. During the predictions phase, the predictions of the classifier are documented in the form of a confusion matrix; listing the number of True Positives (TP), True Negatives (TN), False Positives (FP) and False Negatives (FN). Through these, more meaningful measures may be formulated to highlight the performance of the model in specific scenarios. Sensitivity, specificity and accuracy are among these measures along with the cost of misclassification.  

Accuracy is a fundamental metric used to assess the performance of a model. It quantifies the proportion of correct predictions made by the model relative to the total number of examples evaluated.

\begin{equation}
    \label{eq:accuracy}
    Accuracy = \frac{TP + TN}{TN + FP + FN + TP}
\end{equation}
Another measure used to assess a model's performance is sensitivity, also referred to as \textbf{recall} or True Positive Rate, which determines the proportion of correctly identified real positives.
\begin{equation}
    \label{eq:sensitivity}
    Sensitivity = \frac{TP}{FN + TP}
\end{equation}

The last measure Specificity can be defined as the model's ability to predict a true negative of each category available.
\begin{equation}
    \label{eq:specificity}
    Specificity = \frac{TN}{TN + FP}
\end{equation}
\begin{table*}[!htb]
    \begin{tabular*}{\textwidth}{@{\extracolsep{\fill}}lcccc}
        \toprule
        \textbf{} & \textbf{Precision} & \textbf{Recall} & \textbf{f1-score} & \textbf{support}  \\ \midrule
         \textbf{Likely-reputable} & 0.9992 & 0.9992 & 0.9992 & 1184  \\ 
        \textbf{Illicit} & 0.9970 & 0.9994 & 0.9982 & 663   \\  
        \textbf{accuracy} & & & 0.9984 & 1847 \\ 
        \textbf{macro average} & 0.9981 & 0.9993 & 0.9987 & 1847    \\ 
        \textbf{weighted average} & 0.9988 & 0.9988 & 0.9988 & 1847    \\ 
        \bottomrule
    \end{tabular*}
    \caption{Classification Report of LightGBM model}
      \label{table:regr_report}   
\end{table*}

Given the risk attributed to the incorrect classification of accounts as well as its ability to encompass all criteria surrounding the model’s performance, we deem the AUC as the preferred measurement for identifying the optimal model of LightGBM.

\subsection{Feature Importance}
\label{feat_imp_sect}
LightGBM provides two primary metrics to quantify and interpret feature importance within the trained model: (i) Gain, which represents the cumulative improvement in accuracy contributed by all splits using a given feature, and (ii) Split, which denotes the number of times a feature was used for splitting during tree construction.

To evaluate the model's stability and reliability, its performance is analysed across different train-test size configurations. Feature importance is assessed using both metrics across optimal folds, where some variation in ranking is expected. However, a robust and stable model should exhibit a consistent ranking of key features. The analysis focuses on the top 10 features, which are considered the most influential in determining classification outcomes.
\section{Evaluation}
\begin{figure}[H]
    \centering
    \includegraphics[width=0.5\textwidth]{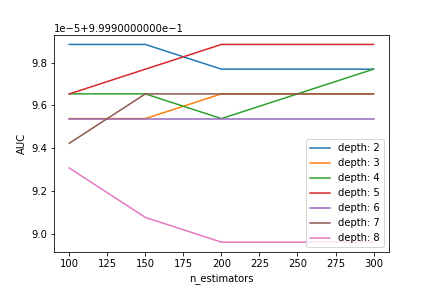}
    \caption[n\_estimators vs max\_depth parameters]{Evaluation of n\_estimators and max\_depth parameters using Grid-Search and 10-fold cross-validation.\label{fig:n_estimators_vs_depth}}
\end{figure}

In order to determine the optimal values for our key parameters, namely the learning rate, number of trees (n\_estimators), and maximum tree depth (max\_depth), grid search cross-validation was employed. Additionally, the performance of the model was evaluated using the AUC measure to identify the best model performance based on the given parameter values. The optimal parameters shown in Fig. \ref{fig:n_estimators_vs_depth} and highlighted by the blue line were identified to be a max depth of 2 and n\_estimators 100 which led to a an AUC of 0.999. The corresponding accuracy for these optimal parameters was 0.9984. 

Table \ref{table:regr_report} shows the classification report of the model with the determined optimal parameters. 

\begin{figure}[htbp]
    \centering
    \includegraphics[width=0.5\textwidth]{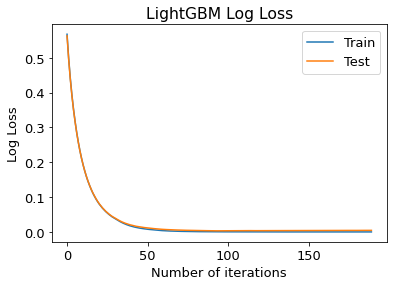}
    \caption[Average logarithmic loss]{The average logarithmic loss, evaluated through 10-fold cross-validation, in relation to the number of iterations of LightGBM.\label{fig:lgbm_log_loss}}
\end{figure}

\begin{figure}
    \centering
    \includegraphics[width=0.5\textwidth]{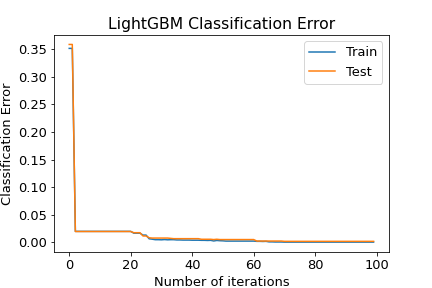}
    \caption[Average classification error]{The average classification error, evaluated through 10-fold cross-validation, in relation to the number of iterations of LightGBM.\label{fig:lgbm_error_loss}}
\end{figure}

In order to gain deeper insights into the learning capabilities of the model, the logarithmic loss and classification error in relation to the number of iterations performed by LightGBM are presented in Fig. \ref{fig:lgbm_log_loss} and Fig. \ref{fig:lgbm_error_loss} respectively.
The logarithmic loss function measures the proximity between the predicted probability and the true value. Additionally, maximising this likelihood is equivalent to minimising the mean square error. Based on the illustration, it can be observed that the learning algorithm begins to converge around 100 iterations.

Finally, through the confusion matrix obtained from the execution of 10-fold cross-validation, the False Positives (Type-2 errors) and False Negatives (Type-1 errors) were identified and are presented in Table \ref{table:fp_fn}. These represent the projects the model misclassified as likely-reputable instead of illicit or illicit instead of likely-reputable, respectively.  In retrospect, although not ideal, we are mostly interested in ensuring that illicit accounts are detected; therefore the ‘False Negatives’.

Comparing the two feature importance parameters discussed in Section \ref{feat_imp_sect}, the 'Split' parameter provided similar feature importance order, especially for the top 10 features. The features maintained similar order, moving 1 places up or down. 

\begin{figure*}[htbp]
    \centering
    \includegraphics[width=0.6\textwidth]{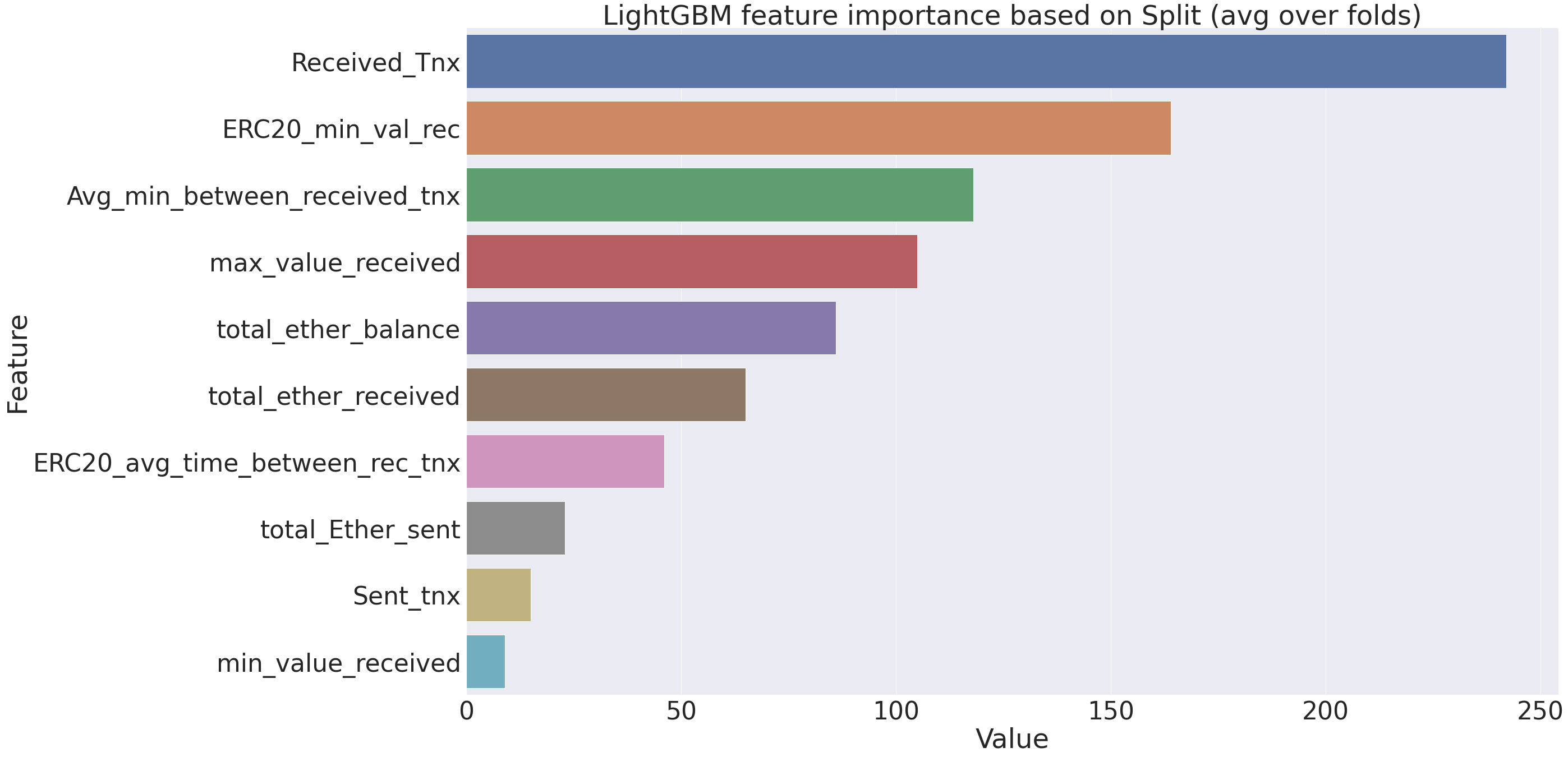}
    \caption[Feature Importance]{Top 10 Features ranking determined by LightGBM based on the 'Split' paramter.\label{fig:feat_imp_split}}
\end{figure*}

The top features pertaining to 'Received no of transactions', 'Minimum ERC20 value received', and 'Average time difference between received transactions (in mins)' held the largest frequency score.  This was followed by 'Minimum value received' and 'Total Ether Balance' (Fig. \ref{fig:feat_imp_split}).

\begin{table*}[!htb]
\begin{tabular}{cc}
        \toprule
        \textbf{False Positives} & \textbf{False Negatives} \\ \midrule
          0xaa602de53347579f86b996d2add74bb6f79462b2 & 0x311f71389e3de68f7b2097ad02c6ad7b2dde4c71   \\ 
         0xc56c2b7e71b54d38aab6d52e94a04cbfa8f604fa &   \\  
        \bottomrule
    \end{tabular}
    \caption{False Positives and Negaives}
      \label{table:fp_fn}   
\end{table*}

\subsection{Misclassified projects}
\subsubsection{False Positives (Type-2 Errors)}
\begin{itemize}
    \item \textbf{Address} 0xaa602de53347579f86b996d2add74bb6f79462b2 \\
    This Ethereum project corresponds to Zipmex Token\footnote{\url{https://etherscan.io/token/0xaa602de53347579f86b996d2add74bb6f79462b2}}, a regulated digital Assets Platform where users can buy, sell, and earn interest on digital assets. Though classified as being 'illicit', it is marked as reputable by CoinGecko. However, from Etherscan, it can be seen that the account's balance has always been 0 ETH and the only transactions the account is involved in are incoming token transactions. Moreover, the project was inactive from January 2020 till January 2022 conducting only 7 transactions. Whilst it seems to be the case that the project is indeed illicit, in the opinion of the author, such minimal activity should indeed not be listed as ``likely-reputable", since there is minimal transaction information to base a project's reputability off of. Therefore, this false-positive has a valid reason for being marked so. 
    
    \item \textbf{Address} 0xc56c2b7e71b54d38aab6d52e94a04cbfa8f604fa \\
    This Ethereum project corresponds to ZUSD Token\footnote{\url{https://etherscan.io/token/0xc56c2b7e71b54d38aab6d52e94a04cbfa8f604fa}}. Similar to Zipmex, the token's balance has maintained a constant value of 0 ETH from 11/12/2019 to 16/05/2023. Furthermore, the account seems to be involved in regular ongoing and incoming transactions however, the maximum number of transactions conducted in any month is 17. Finally, the account is involved in no outgoing transactions. As discussed above, in the opinion of the author, maintaining a zero balance in an account is often seen to demonstrate a project's lack of faith in the project itself and though the the project is listed as 'likely-reputable', this false-positive is reasonably predicted as 'illicit'. 
\end{itemize}

\subsubsection{False Negatives - Type 2}
\begin{itemize}
    \item \textbf{Address} 0x311f71389e3de68f7b2097ad02c6ad7b2dde4c71 \\
    This project was misclassified to be likely-reputable even though it is illicit. The project is marked as Ponzi \footnote{\url{https://www.investor.gov/protect-your-investments/fraud/types-fraud/ponzi-scheme}} on Etherscan. The account was involved in 84,181 transactions with the last transaction being made on 2023/03/051. Moreover, the account had a peak Ether balance of 7,694.31 ETH. Similar to most likely-reputable projects, this account address was not involved in any outgoing transactions and only incoming transactions, with all of these transactions being involved with other unique contract addresses perhaps to only conduct illicit activity with these unique addresses. In the opinion of the author, illicit actors tend to drain balances from such accounts and therefore, the fact that this account is leaving the funds in the account may be a reason why it is incorrectly marked as likely-reputable. \\
\end{itemize}

\section{Conclusions and Future Work}
Assessing the reputability of blockchain projects is critical for investor confidence, providing transparency into the trustworthiness of a given initiative. While this study focuses on identifying reputable projects, prior research has largely approached the issue from the perspective of detecting illicit behavior within blockchain networks. For example, \cite{yu2021ponzi} investigate the identification of Ponzi schemes embedded within smart contracts, while other studies focus on detecting anomalous transactions linked to illegal activities and uncovering money laundering schemes \cite{farrugia2020detection}. While these efforts contribute to mitigating fraudulent behavior, they do not directly address the broader question of evaluating project credibility.

While these approaches contribute significantly to the detection of illicit activities within blockchain networks, they remain focused on identifying fraudulent behavior rather than evaluating a project's overall reputability. They do not offer a comprehensive assessment of project credibility, particularly at the level of individual accounts, nor do they provide the precision required for determining reputability with a high degree of confidence.

By integrating diverse concepts and methodologies from related fields, particularly feature extraction and feature importance, we propose an innovative approach to assess the reputability of projects on the Ethereum network at an individual account level. This research utilises the advanced LightGBM classification model. By employing a well-balanced dataset consisting of both "illicit" and "likely-reputable" accounts, the model demonstrated exceptional effectiveness, yielding an average AUC value of 0.999 and a standard deviation of 0.0003 across 10-fold cross-validation. It is worth noting that the model exhibited robust generalisation capabilities and displayed resilience against overfitting. This observation is supported by the low standard deviations observed in both AUC and accuracy, as well as the consistent performance achieved even with a small number of k-folds during cross-validation. Despite efficiency not being the primary focus, the minimal resources utilised underscore the scalability of the system. 

When examining the rank of feature importance, the average time difference between received transactions is the most important factor in account classification. Likely-reputable accounts have an average value of 165.5 seconds, while illicit accounts have an average of 47 minutes. This suggests that likely-reputable accounts receive multiple incoming transactions from unique addresses, highlighting a significant difference between illicit and likely-reputable accounts.

\subsection{Future work}
This study establishes a foundational framework for future research in the domain of blockchain project evaluation. One potential avenue for further exploration is assessing the applicability of the proposed approach across different blockchain networks. Given that some projects operate across multiple chains, deploying tokens and smart contracts on distinct platforms, developing cross-blockchain models capable of identifying both illicit and likely reputable projects would be a valuable extension.

Another promising direction involves refining the methodology to assess reputability at a more granular transactional level. Adapting the framework to evaluate individual transactions rather than entire projects could enhance precision and provide deeper insights into patterns of trustworthy and fraudulent behavior within blockchain ecosystems.

An alternative approach to assessing reputability involves analysing Ethereum project smart contract code to identify common structural and functional characteristics among reputable projects. By extracting and modeling these shared features, a more refined classification framework can be developed, incorporating additional reputability indicators for improved accuracy. Furthermore, extending the analysis beyond standard Ethereum transactions to include internal transactions—those executed within smart contracts—could provide deeper insights. Integrating this additional layer of data into the LightGBM model may enhance its classification capabilities, improving the distinction between reputable and potentially fraudulent projects.

Another potential enhancement involves incorporating graph analysis to refine feature extraction.


\bibliographystyle{ACM-Reference-Format}
\bibliography{sample-base}

\end{document}